\documentclass[annual]{acmsiggraph}

\TOGonlineid{45678}
\TOGvolume{0}
\TOGnumber{0}
\TOGarticleDOI{1111111.2222222}
\TOGprojectURL{}
\TOGvideoURL{}
\TOGdataURL{}
\TOGcodeURL{}

\usepackage{amsthm}
\usepackage{algorithmic}
\usepackage{algorithm}

\theoremstyle{plain}
\newtheorem{thm}{Theorem}[section] 

\theoremstyle{definition}
\newtheorem{defn}[thm]{Definition} 

\theoremstyle{proposition}

\title{As-exact-as-possible repair of unprintable STL files}

\author{Marco Attene\thanks{\emph{This work is partly supported by the EU H2020-FoF-2014-2015/H2020-FoF-2015 Project N. 680448 (CAxMan) and by the international joint project on "Mesh Repairing for 3D Printing Applications" funded by Software Architects Inc (WA, USA). Thanks are due to colleagues at IMATI for helpful discussions.} Author's e-mail: marco.attene@ge.imati.cnr.it}\\CNR-IMATI-GE}
\pdfauthor{Marco Attene}


\begin{document}


\maketitle

\begin{abstract}


\textbf{Purpose}: The class of models that can be represented by STL files is larger than the class of models that can be printed using additive manufacturing technologies. Stated differently, there exist well-formed STL files that cannot be printed.
In this paper such a gap is formalized and a fully automatic procedure is described to turn \emph{any} such file into a printable model.

\textbf{Approach}: Based on well-established concepts from combinatorial topology, we provide an unambiguous description of all the mathematical entities involved in the modeling-printing pipeline. 
Specifically, we formally define the conditions that an STL file must satisfy to be printable and, based on these, we design an as-exact-as-possible repairing algorithm.

\textbf{Findings}: We have found that, in order to cope with all the possible triangle configurations, the algorithm must distinguish between triangles that bound solid parts and triangles that constitute zero-thickness sheets. Only the former set can be fixed without distortion.

\textbf{Originality}: Previous methods that are guaranteed to fix all the possible configurations provide only approximate solutions with an unnecessary distortion. Conversely, our procedure is as exact as possible, meaning that no visible distortion is introduced unless it is strictly imposed by limitations of the printing device. Thanks to such an unprecedented flexibility and accuracy, this algorithm is expected to significantly simplify the modeling-printing process, in particular within the continuously emerging non-professional "maker" communities.

\end{abstract}


\keywordlist


\copyrightspace

\section{Introduction}

Today fabricating an appropriate 3D model using a low-cost 3D printer is nearly as easy as printing a textual document, but creating a 3D model which is actually "appropriate" for printing is definitely complicated. Many STL files have defects that make them unsuitable for printing (e.g. self-intersections, zero-thickness walls, incomplete geometry, ...), and the computation of a valid toolpath becomes complicated and ill-posed when the mesh does not enclose a polyhedron in an unambiguous manner.

Note that a model can appear perfectly fine when visualized, but the presence of \emph{hidden} defects may prevent the possibility to print it. Existing mesh repairing algorithms allow expert users to fix virtually any sort of defect, provided that the user is able to accommodate the requirements on the input that most such algorithms have \cite{attene2013}. Unfortunately the problem is much more complex if one considers the so-called "maker movement". This quickly-growing community is mostly made by amateurs and persons whose background is not technical enough to recognize all the defects, to select the most appropriate repairing algorithms, and to stack them into workflows where each step guarantees the right working conditions for the next step. These users would prefer a single repairing method which is completely automatic and does not have any specific requirement on the input. Key actors such as Autodesk and Microsoft released popular methods to automatically repair STL files (Section \ref{sec:sota}), but their solutions are not guaranteed to succeed and, when they do, the accuracy of the fixed model can be suboptimal.

This paper provides a rigorous formalization of the mathematical entities involved in the modeling-printing pipeline and, based on it, describes an algorithm to turn any STL file into a printable model. No assumption is made on the input STL file, the user is not forced to interact with the algorithm, and no visible distortion is introduced if it is not strictly required by the specific printing technology.
To the best of our knowledge no previous mesh repairing algorithm encapsulates all of these characteristics. A prototype implementation has been developed, and the results reported in section \ref{sec:results} show that it could accurately fix STL files that could not be properly repaired by state-of-the-art algorithms.

\section{State of the art}
\label{sec:sota}

\begin{figure*}[!tb]
    \centering
  \includegraphics[width=.99\linewidth]{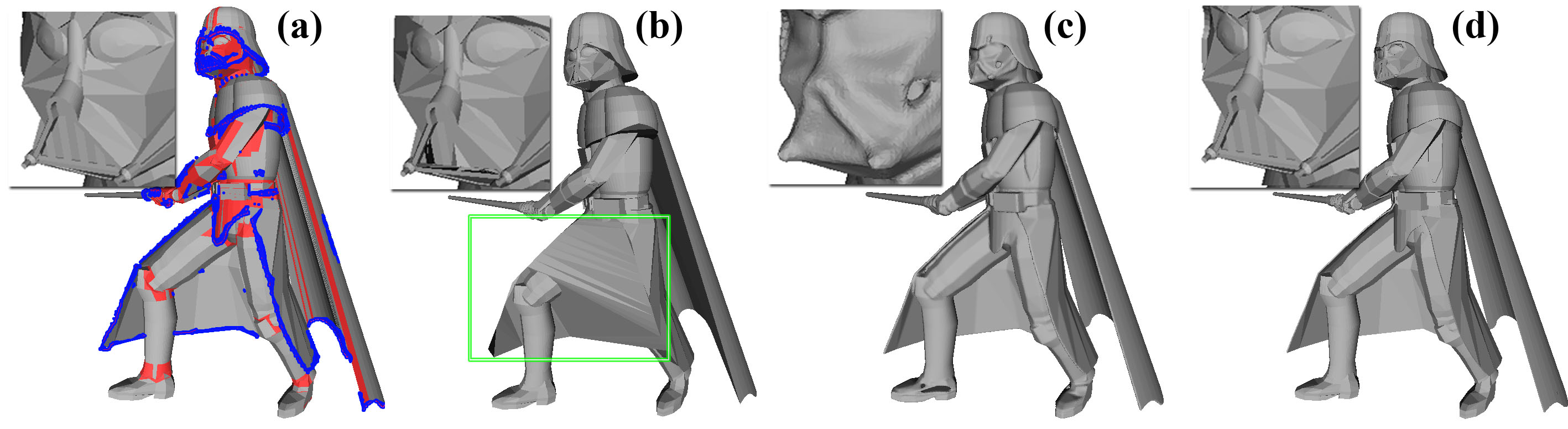}
  \caption{\label{fig:darth}From left to right: an original STL file (a) with self-intersections (in red) and open surfaces used to model sheet-like features (e.g. helmet and cloak, boundaries in blue), and three repaired versions produced by NetFABB cloud service (b), Autodesk Meshmixer (c), and our algorithm (d). The hole-filling employed by NetFABB produces a coarse result, especially in the cloak (e.g. within the green box), whereas MeshMixer evenly distorts the whole model.}
\end{figure*}

Some repairing methods transform the input into an intermediate volumetric representation and construct a new mesh out of it \cite{taoju04} \cite{kobbelttog} \cite{chenwang13}. In a new trend of methods, a 3D mesh is converted into an implicit representation, and all the subsequent operations (including the slicing) are performed on this representation \cite{huang13} \cite{huang14}. These methods are very robust but necessarily introduce a distortion.
When distortion is not allowed and self-intersections must be removed \cite{attene2014}, some approaches rely on exact arithmetic, while some others can losslessly convert the input into a finite precision plane-based representation, and then reconstruct a provably good fixed mesh out of it \cite{campen10} \cite{wang13}.
The aforementioned approaches are useful only if the input actually encloses a solid, while they are not really suitable to fix open meshes such as the example in Figure \ref{fig:darth}.
For a more comprehensive overview of mesh repairing methods, we point the reader to \cite{attene2013}.

STL files with particular defects make the most diffused slicing software (e.g. Slic3r, Skeinforge, ...) either fail or produce machine instructions that lead to printing failures. Some slicers attempt to repair the model on a per-slice basis (e.g. by closing small gaps in open curves), but complex ambiguities cannot be dealt with. Thus, repairing these meshes before the slicing is crucial, but robust repairing methods are either approximate \cite{yau2003}\cite{taoju04} or make strong assumptions on the \emph{solidity} of the input \cite{campen10}. We currently miss a method that can cope with \textbf{any} STL file while preserving all the visible surfaces with no distortion. 
If the input mesh has open boundaries, popular web-based mesh fixing services (https://netfabb.azurewebsites.net/) treat these boundaries as holes and patch them with new triangles; unfortunately, in many cases this produces quite coarse results (see Figure \ref{fig:darth}(b) and Figure \ref{fig:hagia}(g)). Furthermore, some STL files might have no open boundaries though being not solid. To produce such a model, for example, one may just take a triangulated cube and set all the Z coordinates to zero: on such a flat cube, NetFABB cloud service produces an empty output.
In Autodesk's Meshmixer (http://www.meshmixer.com) the input STL can be successfully fixed even if it has open boundaries but at the cost of an overall approximation (Figure \ref{fig:darth}(c)).
Blender's 3D print toolbox (http://www.blender.org) includes tools to fix common issues in STL files; unfortunately, intersecting meshes are only loosely treated by removing vertices that remain enclosed in the resulting solid, whereas a precise remeshing is not provided. Similarly, the only way to fix these models in MeshLAB (http://www.meshlab.net/) is to select and remove the self-intersecting facets, and then to fill the resulting surface holes with new triangles. This is the same approach employed by other popular tools such as FreeCAD's \emph{Evaluate and Repair Mesh} tool (https://www.freecadweb.org/) and MeshFIX (http://meshfix.sourceforge.net/) that repeatedly remove intersections and patch holes up to convergence; as stated by the authors, however, these tools are necessarily approximated and provide acceptable solutions only for raw digitized meshes. Differently from the algorithm introduced in this paper, none of the aforementioned tools is guaranteed to produce a printable result for \emph{any} possible input STL file unless an unnecessary distortion is allowed.

While building on existing techniques, this article introduces an important original approach: to convert \textbf{any} set of triangles into a solid and printable model, one may distinguish between visible volumes and sheets, and treat them separately. Thanks to such a separation, we can define what \emph{as-exact-as-possible} repairing means.
Our original contributions cover both formal and practical aspects: definitions are given to classify new geometric concepts, including an exact definition of \emph{printable geometry}, whereas a new robust and as-exact-as-possible algorithm is described to turn any STL file into a set of printable volumes.

\section{Terminologies and Definitions}
\label{sec:terminology}
If an STL file needs no repairing, i.e. it represents a well-defined closed volume, any slicer can easily calculate a toolpath that follows the exact intersection of a set of planes with the surface. This solution, however, is suboptimal because it does not consider the thickness of the extruded material, and the resulting physical object would be slightly larger than the virtual model. Conversely, an optimal slicer should consider such a thickness and produce a toolpath that follows a slightly negative offset of the STL surface.
If the set of triangles has zero thickness (e.g. the cloak in Figure \ref{fig:darth}), an optimal slicer cannot do better than following the surface, and the printed object would be inevitably thicker than the model. Thus, in this case we are allowed to slightly thicken the STL model without causing any additional distortion because the distortion would be there independently of our intervention.
On these premises, we say that a mesh repairing algorithm is \emph{as-exact-as-possible} if it does not cause any additional distortion with respect to the one that an optimal slicing would introduce in the printed model.

When dealing with STL files and meshes, various authors in academic literature use similar terms to indicate slightly different concepts. For example, while some papers deal with \emph{self-intersecting} triangle meshes, some others define a triangle mesh as a (Euclidean) simplicial complex, even if a simplicial complex cannot have self-intersections by definition. Since in the scope of this article there is no room for similar ambiguities, the remainder of this section formalizes a series of concepts that constitute the building blocks for the main definition of \emph{printability}. This definition is then used to define a conversion algorithm that produces printable meshes out of raw triangle collections.
Specifically, the following definitions conceptualize configurations that may lead to failures along the printing pipeline, and are used to progressively restrict the class of \emph{valid} models. We start with all the models that can be represented by an STL file (Sec. \ref{sec:stl_models}) and put them in relation with the notion of triangle mesh (Sec. \ref{sec:meshes}). Then, Sec. \ref{sec:polyhedra} characterizes the subset of meshes that do not self-intersect, and Sec. \ref{sec:printability} further specializes the subset of non-intersecting meshes that enclose printable solids.

In the remainder we assume that the only reliable information brought by an STL file is the vertex position. Hence, our formalization intentionally disregards any additional information such as, e.g., triangle normals or vertex orientation.
We also assume that the reader is familiar with the fundamental concepts of combinatorial topology. In particular, we make use of terms such as, e.g., \emph{abstract simplicial complex} and \emph{face} of a simplex.
Roughly speaking, a 2-dimensional simplicial complex is a set of simplexes (i.e. vertices, edges and triangles) that connect \emph{properly}, meaning that the intersection of any two simplexes is either empty or it is a $face$ of both. Any edge or vertex $\sigma$ of a triangle $t$ is said to be a $face$ of $t$, and any vertex $v$ of an edge $e$ is a face of $e$. Simplicial complexes can be either $Euclidean$ (each vertex corresponds to a point in space) or $abstract$ (vertices are unspecified $abstract$ entities and simplexes are just sets of these abstract vertices). Hence, an abstract simplicial complex may be used to represent the pure connectivity of an Euclidean simplicial complex while disregarding its geometry. When no qualifier is used, a simplicial complex is meant to be Euclidean.
Formal definitions can be found in \cite{glaser}, \cite{stillwell93}, and \cite{ferrario11}.

\subsection{STL Models}
\label{sec:stl_models}
An STL file is essentially an unstructured collection of triangles.
A triplet of \emph{vertices} $<v_1, v_2, v_3>$ with $v_i \in R^3$ is an \emph{STL triangle}.
An STL triangle is \emph{regular} if it has three different vertices.
Two STL triangles are \emph{equivalent} if they have the same vertices up to permutations (i.e. $<v_a, v_b, v_c>$ and $<v_b, v_a, v_c>$). Note that two triangles can be equivalent even if their vertex ordering are opposite.

\begin{defn} \emph{STL Model - } A finite collection of STL triangles $T = \{t_i = <v_{i1}, v_{i2}, v_{i3}>, v_{ij} \in R^3\}$ is an STL model (or simply an STL), and the subset of $R^3$ made by the union of all the $v_{ij}$'s is its \emph{vertex set} and is denoted by $V(T)$. Example: $T = \{<<1, 0, 0>, <0, 1, 0>, <0, 0, 0>>, <<0, 1, 0>, <1, 0, 0>, <1, 1, 0>>\}$ and $V(T) = \{<1, 0, 0>, <0, 1, 0>, <0, 0, 0>, <1, 1, 0>\}$. $T$ is \emph{regular} if all its triangles are regular. Any maximal subset of non-equivalent triangles in $T$ is called a \emph{Layer} of $T$ and is denoted by $L(T)$. Note that $V(L(T)) = V(T)$.\end{defn}

\subsection{Meshes}
\label{sec:meshes}
A triangle mesh is a set of triangles with an explicit structure.

\begin{defn} \emph{Triangle Mesh - } A triangle mesh is a pair $M = (V, S)$, where $V$ is a set of points in $R^3$, $S$ is a pure two-dimensional abstract simplicial complex \cite{glaser}, and there is a bijective map between the set of vertices of $S$ and $V$. Such a map is called \emph{vertex embedding} and is denoted as $\varphi: S_0 \rightarrow V$, where $S_0$ denotes the set of vertices of $S$.
Thus, any 0-simplex $s$ in $S$ is mapped to exactly one of the points in $V$ through the vertex embedding.
Notice that $S$ is not necessarily a combinatorial manifold. Since this article does not deal with non-triangular meshes, in the remainder we shall omit the qualifying term "triangle".\end{defn}

A mesh $M = (V, S)$ is \emph{closed} if S has no boundary, that is, if any 1-simplex in $S$ is a face of an even number of 2-simplexes in $S$ \cite{stillwell93}.

\begin{defn} \emph{STL-induced Mesh - } Let $T$ be a regular STL and let $L(T)$ be one of its layers. Also, let $M = (V(T), S(T))$ be a mesh where $V(T)$ is the vertex set of $T$ and $S(T)$ is obtained by considering each triangle $t_i$ in $L(T)$ as an abstract 2-simplex whose vertices are mapped to $t_i$'s vertices. $M$ is called a $T$-induced mesh.\end{defn}

\begin{defn} \emph{Geometric Realization - } Let $M = (V,S)$ be a mesh. The geometric realization of a k-simplex $s = \{v_0, ..., v_k\} \in S$ is the convex hull of the points ${\varphi(v_0), ..., \varphi(v_k)}$ and is denoted by $|s|$. Thus, any point in $|s|$ can be expressed as $\lambda_0 \varphi(v_0) + ... + \lambda_k \varphi(v_k)$, $1 \geq \lambda_i \geq 0$, and $\lambda_1 + ... + \lambda_k = 1$.
The union of the geometric realization of all the simplexes is the geometric realization of the mesh, denoted by $|M|$.\end{defn}

\begin{defn} \emph{Degenerate Realization - } Let $M = (V,S)$ be a mesh. The geometric realization of a 2-simplex in $S$ is a properly 2-dimensional subset of $R^3$ only if its three vertices are mapped to points in $V$ which are in general position (i.e. the three points are neither coincident nor collinear). In all the other cases the geometric realization is said to be \emph{degenerate}.\end{defn}

\begin{defn} \emph{Self-intersection - } Let $M = (V,S)$ be a mesh and let $s_1$ and $s_2$ be two simplexes in $S$. Also, let $X$ be the intersection of $|s_1|$ and $|s_2|$. If $X$ is not empty and $X$ is not the geometric realization of a simplex in $S$ which is a face of both $s_1$ and $s_2$, then $X$ is called a self-intersection of $M$.\end{defn}

\subsection{Polyhedra}
\label{sec:polyhedra}
\begin{defn}
\emph{Polyhedron - } 
A polyhedron is the geometric realization of a mesh $M$ without self-intersections. In this case the geometric realizations of the simplexes in $M$ are Euclidean simplexes forming an Euclidean simplicial complex \cite{ferrario11}.
Since any 2-simplex with a degenerate realization leads to self-intersections, a mesh with degenerate elements does not admit a polyhedron.
We observe that a polyhedron is not necessarily a two-manifold with boundary, and say that a polyhedron $|M|$ is closed if the non-self-intersecting mesh $M$ is also closed.\end{defn}

The concept of \emph{outer hull} is central in this work, therefore it is worth providing a more accessible intuitive introduction of such an object before giving its formal definition. Roughly speaking, the outer hull of a polyhedron is the subset of its points which can be reached from infinity through continuous paths. In other words, we may loosely say that the outer hull is made by the points of the polyhedron which are visible from the outside. Furthermore, if a point of the outer hull can be reached from one side only, then it is part of the so-called \emph{solid outer hull}.

\begin{defn}
\label{def:outerhull}
\emph{Outer Hull - } Let $|M|$ be a polyhedron and let $p \in R^3$ be a point out of it (e.g. if $v$ is the vertex in $V$ having the maximum $x$ coordinate, take $p = v+<1, 0, 0>$). Let $|M'| \subseteq |M|$ be the set of points $q$ for which there exists a path $<q, p>$ which does not contain any other point of $|M|$ (excluding $q$ itself). Then $|M'|$ is the outer hull of $|M|$.
The outer hull is a polyhedron.\end{defn}

\begin{defn}
\label{def:solidouterhull}
\emph{Solid Outer Hull - } Let $|M|$ be a polyhedron and let $|M'|$ be its outer hull. 
With reference to the above definition \ref{def:outerhull}, let $O$ be the union of $|M'|$ and all the points in space which are path-connected with $p$ on paths that do not contain points of $|M|$. The Solid Outer Hull $M_{solid}$ of $M$ is the boundary of $O$.\end{defn}

\subsection{Printability Condition}
\label{sec:printability}
\begin{figure}[h]
    \centering
  \includegraphics[width=.99\linewidth]{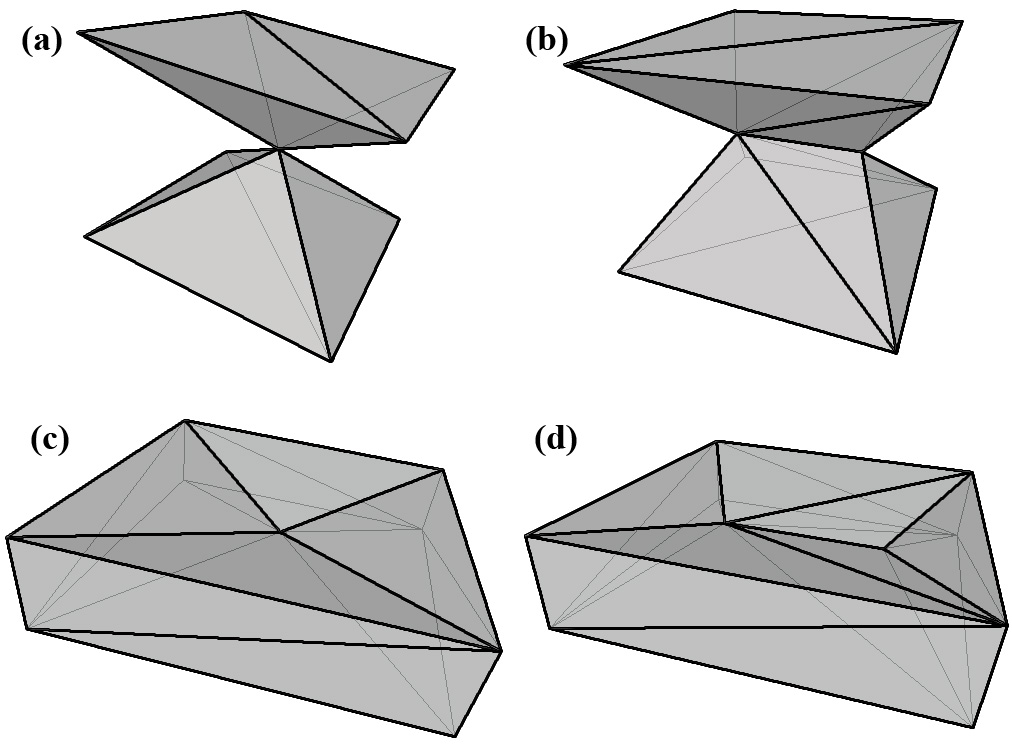}
  \caption{\label{fig:nmconfigs}All these meshes have singularities that make them non manifold. (c) can be obtained from a cylinder by pushing the central points of its two bases towards the same point in the center of the cylinder. (d) can be obtained as for (c), but instead of pushing two points we push two edges towards a single common edge. From a topological perspective, (c) and (d) can be seen as the complementary of (a) and (b) respectively, but only (c) and (d) are manifold-connected.}
\end{figure}

We characterize STL files that unambiguously represent a single \emph{sufficiently connected} solid object. Intuitively, two pyramids that touch at their apex only are not sufficiently connected (i.e. the small connection is too weak to keep the two parts together). Though it is tempting to model this class of objects as the class of 2-manifolds, there are cases where singularities occur though the model is solid enough (Figure \ref{fig:nmconfigs} (c) and (d)). Notice that these cases cannot be split in parts to be printed separately as one could do with the aforementioned two pyramids.
This intuitive concept is formalized by the notion of \emph{manifold-connectedness} given in Def. \ref{def:mcpolyhedron}.

Any two-dimensional abstract simplicial complex with singularities (i.e. non-manifold vertices/edges) can be decomposed into a collection of manifold complexes by properly duplicating singular simplexes. This duplication procedure is purely combinatorial and does not consider possible vertex embeddings. The resulting set of combinatorial manifolds is provably unique and is called a \emph{manifold decomposition} of the input complex \cite{deflosgp06}. Without loss of generality, if we assume to deal with a single connected complex with singularities, its manifold decomposition may be either a set of $N>1$ complexes (e.g. Figure \ref{fig:nmconfigs} (a) and (b)) or another single connected complex (e.g. Figure \ref{fig:nmconfigs} (c) and (d)).

\begin{defn}
\label{def:mcpolyhedron}
\emph{Manifold-connected Polyhedron - } Let $|M|$ be a polyhedron. $|M|$ is manifold-connected if the manifold decomposition of $M$ is made of a single component.\end{defn}

All the definitions given up to this point are necessary to give the following definition of printability.

\begin{defn}
\label{def:printable}
\emph{Printable STL - } An STL model $T$ is printable if there exists a $T$-induced mesh whose realization is a closed and manifold-connected polyhedron that coincides with its outer hull.
\end{defn}

\section{Conversion Algorithm}
\label{conversion}
An arbitrary STL model may require a sequence of repairing operations to become printable according to Definition \ref{def:printable}.
Our only assumption is that the STL file is syntactically well-formed: no other requirement is necessary.
The algorithm first turns the STL model into a possibly non-manifold mesh without intersections (Sec. \ref{sec:stltopoly}).
Then, while computing the outer hull $|M|$, the algorithm splits the outer hull itself in two subsets: $M_{solid}$ and $M_{sheet}$, where $M_{solid}$ represents the union of all the parts of $|M|$ that enclose a volume (see Def. \ref{def:solidouterhull}), and $M_{sheet}$ is the union of all the remaining \emph{sheet}-like parts of $|M|$ (Sec. \ref{sec:polytoprint}).
If $M_{sheet}$ is not empty, each component in $M_{sheet}$ is transformed to a thin solid through a thickening procedure (Sec. \ref{sec:easycase}) and the whole process is repeated. Note that $M_{sheet}$ is necessarily empty on the second iteration.
In a last \emph{polishing} phase, each of the printable components in $M_{solid}$ is isolated (Sec. \ref{sec:easycase}) and the algorithm terminates.

\subsection{From STL files to polyhedra}
\label{sec:stltopoly}
After having deleted all the irregular and degenerate triangles we can easily extract a layer by simply removing possible equivalent triangles. The remaining triangles are used to create an induced mesh. If such a mesh has no self-intersections, then it is a valid Euclidean simplicial complex \cite{ferrario11} whose realization is a polyhedron.
If self-intersections occur, new simplexes must be created to represent them in the abstract complex too \cite{attene2014}, so that the result is a possibly non-manifold mesh without intersections.

\subsection{From polyhedra to printable polyhedra}
\label{sec:polytoprint}
Intuitively, the outer hull $|M|$ can be extracted by region growing: a triangle which is known to be on the outer hull is selected, and from such a seed the remaining parts are reached by crossing edges while \emph{walking} on the outer side of the complex.
If the input complex is a single closed 2-manifold, the region reaches all the triangles, and the result is obviously printable according to Def. \ref{def:printable}.
If the complex is a collection of disjoint closed 2-manifolds, the region covers one of the printable components only. Hence, the process must be iterated and, in the end, each of the so-extracted components which is spatially contained in other components is removed. The remaining parts form a collection of printable models corresponding to the outer hull of the input.
If we allow singular vertices, but not singular edges, this procedure can still be employed.

If we allow singular edges and "border" edges (i.e. with only one incident triangle), things become much more complicated because the outer hull might be no longer a collection of printable components and unorientable parts might come into play. Thus, while tracking the outer hull $|M|$, we need to recognize which of its subsets form closed polyhedra ($M_{solid}$) and which other subsets require a thickening ($M_{sheet}$).

To support these cases, we first orient the seed and then propagate such an orientation as the region grows across edges. When a singular edge is encountered, we grow on the triangle which is the "most external" according to the orientation of the triangle $t$ we are coming from: using a metaphor, an ant walking on the outer side of $t$ towards the singular edge would proceed onto such a "most external" triangle. If different propagation directions induce opposite orientations on a same triangle, that triangle is eventually moved to $M_{sheet}$.

If the result is not manifold-connected (e.g. the growing would cover the whole model in Figure \ref{fig:nmconfigs}(b)), we just split it into its manifold-connected parts \cite{deflosgp06}.

The following subsection formalizes the aforementioned intuitive concepts and procedure.

\subsubsection{Non-manifold region growing}
\label{sec:easycase}
We classify each edge based on the number of triangles that share that edge (i.e. its \emph{incident triangles}). Specifically, an edge is: \emph{on boundary} if it has only one incident triangle; \emph{2-connected} if it has exactly two incident triangles; \emph{singular} if it has more than two incident triangles.

\begin{defn} \emph{Triangle Fan - } Let $e$ be an edge, let $\mathcal{T}(e)$ be the set of all its incident triangles, and let $t_0$ be one triangle in $\mathcal{T}(e)$. A \emph{triangle fan} at $e$ from $t_0$ is an ordered list $Fan(e, t_0) = <t_0, t_1, ..., t_n>$ whose elements are all and only the triangles in $\mathcal{T}(e)$, $t_0$ is the first element, and the triangles preserve their radial order around $e$. When $e$ is singular, the pair $(e, t_0)$ admits two triangle fans that can be distinguished by the radial order direction (clockwise and counter-clockwise).\end{defn}

Let us assume that the normal $n$ at $t_0$ is well-defined and reliable, and let $v$ be the vertex of $t_0$ that does not belong to $e$. A triangle fan is \emph{upward} with $t_0$ if $t_0$ is the only element in the fan or if the second element is the first triangle after $t_0$ when turning around $e$ in the direction specified by the vector $n$ applied on $v$. Hence, under the just mentioned assumptions, a pair $(e, t_0)$ has exactly one upward triangle fan and possibly one downward fan (non-upward, if $e$ is singular).

\begin{defn}
\label{def:tricontinuation}
\emph{Continuation - } Let $e$ be an edge, and let $t$ be a triangle of $Fan(e, t_0)$. The \emph{continuation} of $t$ in $Fan$ is the first triangle after $t$ if $t$ is not the last element of the list, or the first element of $Fan$ if $t$ is the last element. Note that if $e$ is on boundary, the continuation of its incident triangle $t$ is $t$ itself.\end{defn}

\begin{defn}
\label{def:edgeconnected}
\emph{Edge-connected Polyhedron - } Let $|M|$ be a polyhedron. $|M|$ is edge-connected if any pair of 2-simplexes in $M$ is edge-connected. Two 2-simplexes $s_a$ and $s_b$ are \emph{edge-connected} if there exists a sequence $s_a=s_1, ..., s_n=s_b$ such that $s_{i-1}$ and $s_i$ intersect at a common 1-simplex \cite{deflosgp06}.\end{defn}

At each iteration, we consider an edge-connected component of the input and track its outer hull by starting from a seed triangle. The seed must be guaranteed to be part of the outer hull and its correct orientation/normal must be known without ambiguity (i.e. we must know exactly on which of its two sides we can find the solid). To ensure this, we select a starting extreme vertex $v_0$ as the one having the maximum $x$ coordinate. We then pick all the edges incident at $v_0$ and select the one (let it be $e_0$) whose normalized vector $n = <n_x, n_y, n_z>$ has the smallest absolute value $n_x$. Finally, we pick all the triangles incident at $e_0$ and select the one (let it be $t_0$) whose normal $n = <n_x, n_y, n_z>$ has the largest absolute value $n_x$. If $n_x$ is negative, we invert the orientation of $t_0$.

With reference to algorithm \ref{al:conncomp}, all the triangles but $t_0$ are initially untagged, while $t_0$ is tagged as $outer$, meaning that its outer side is path-connected with an external point (see Def. \ref{def:outerhull}). $t_0$ constitutes the initial region to be grown, and its three edges $e_1, e_2$ and $e_3$ define the region border. Specifically, the region border is made of three pairs $<t_0, e_1>$, $<t_0, e_2>$, $<t_0, e_3>$. At each step, we pick one of the pairs $<t_i, e_j>$ from the border and extract the continuation of $t_i$ in its upward $Fan$ at $e_j$. Let $t_{new}$ be such a continuation. If $t_{new}$ is consistently oriented with $t_i$, then we tag $t_{new}$ as $outer$ and include it in the region. Otherwise, if it is not consistently oriented, we invert its orientation, tag it as $outer$ and include it in the region: however, when inverting the orientation we check whether $t_{new}$ was already tagged as $outer$ and, if so, we switch its old tag to $inner$. As a consequence, after its inclusion in the region $t_{new}$ has both an $inner$ and an $outer$ tag. For simplicity, we say that such a triangle has a $doublesided$ tag. We remind that two adjacent triangles are consistently oriented if their vertices induce opposite orientations on the common edge. If $e_j$ is on the boundary, $t_{new}$ and $t_i$ are the same triangle and, therefore, induce the same orientation on $e_j$. As a consequence, any boundary edge reached by the region growing produces a $doublesided$ tag for its incident triangle.

\begin{figure}[h]
    \centering
  \includegraphics[width=.99\linewidth]{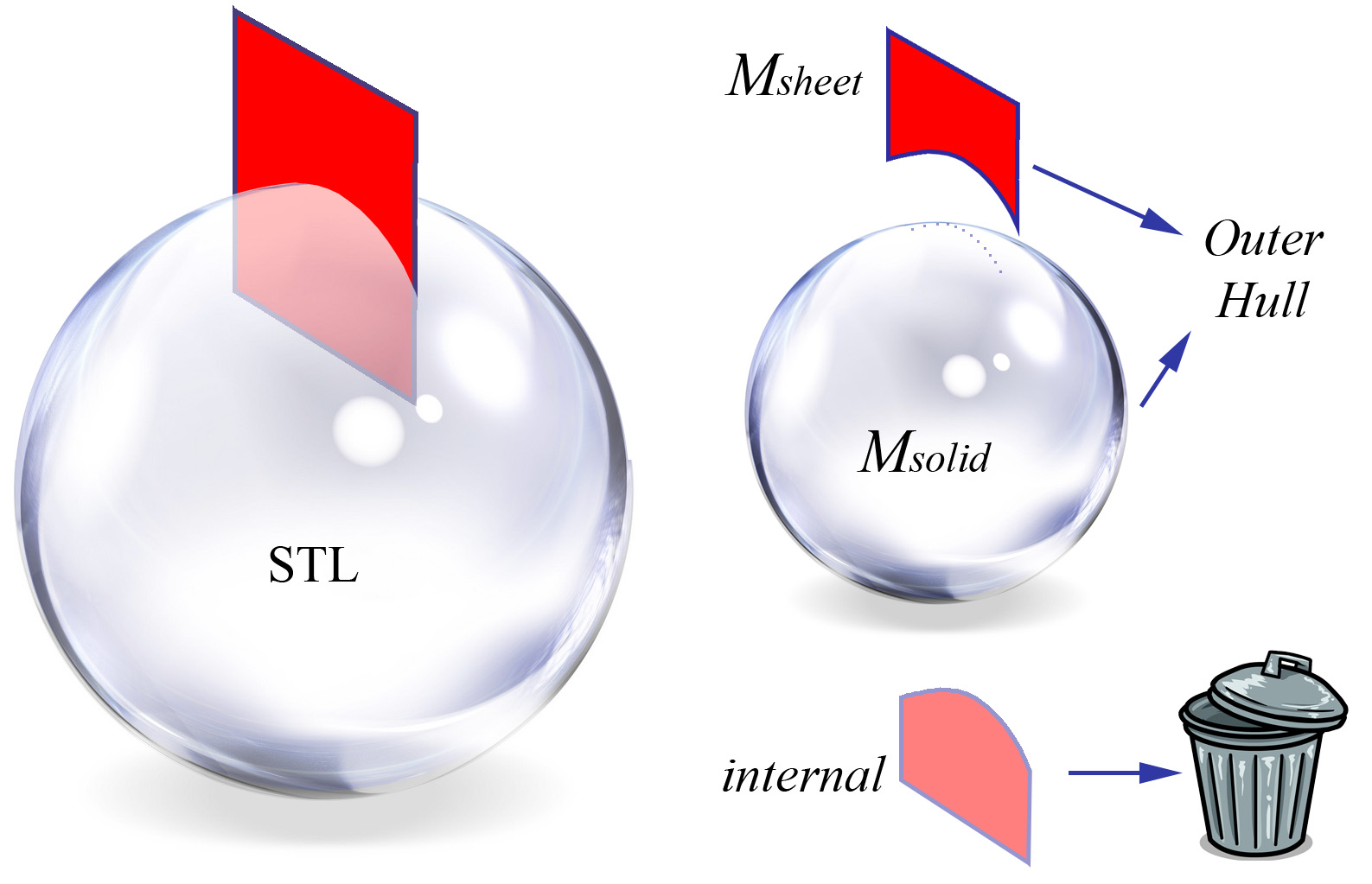}
  \caption{\label{fig:solidsheetinternal}A triangle can be part of $M_{solid}$, part of $M_{sheet}$, or $internal$. Internal triangles are deleted, whereas the other two sets form the outer hull.}
\end{figure}

At the end of the process, a triangle can be untagged or tagged as either $outer$ or $doublesided$. $outer$ triangles constitute the solid outer hull of the edge-connected component at hand (i.e. its $M_{solid}$), whereas the $doublesided$ triangles constitute its $M_{sheet}$. The remaining $untraversed$ triangles are edge-connected with the seed but cannot be reached from infinity, which means they are not part of the outer hull (Figure \ref{fig:solidsheetinternal}).

Based on the just defined concepts, our procedure can be summarized by Algorithm \ref{al:polydec} that, in turn, runs Algorithm \ref{al:conncomp} for each of the edge-connected components.

\begin{algorithm}[!ht]

\caption{The outer hull extraction algorithm for a single edge-connected component}
\label{al:conncomp}
\begin{algorithmic}[1]

\REQUIRE An edge-connected polyhedron $|M|$
\REQUIRE A correctly oriented seed triangle $t_0$ in $M$
\ENSURE $M_{solid}$ and $M_{sheet}$

\STATE $e_1, e_2, e_3$ = the three edges of $t_0$
\STATE List $B = {<t_0, e_1>, <t_0, e_2>, <t_0, e_3>}$
\STATE tag $t_0$ as $outer$

\WHILE {$B$ is not empty}
	\STATE $<t_i, e_j>$ = first element of $B$
	\STATE remove $<t_i, e_j>$ from $B$
	\STATE $t_{new}$ = upward continuation of $t_i$ at $e_j$
	
	\IF {$t_{new}$ is not tagged as both $outer$ and $inner$}
		\IF {$t_{new}$ is consistently oriented with $t_i$}
			\IF {$t_{new}$ is not tagged as $outer$}
				\STATE tag $t_{new}$ as $outer$ and update $B$
			\ENDIF
		\ELSE
			\STATE invert $t_{new}$'s orientation
			\IF {$t_{new}$ is tagged as $outer$}
				\STATE tag $t_{new}$ as $inner$
			\ELSE
				\IF {$t_{new}$ is tagged as $inner$}
					\STATE tag $t_{new}$ as $outer$
				\ENDIF
			\ENDIF
		
			\IF {$t_{new}$ is not tagged as $outer$}
				\STATE tag $t_{new}$ as outer and update $B$
			\ENDIF
		\ENDIF
	\ENDIF
\ENDWHILE

\FORALL {$t_i$ in $M$}
	\IF {$t_i$ is tagged as both $outer$ and $inner$}
		\STATE move $t_i$ from $M$ to $M_{sheet}$
	\ELSE
		\IF {$t_i$ is tagged as $outer$}
			\STATE move $t_i$ from $M$ to $M_{solid}$
		\ELSE
			\STATE delete $t_i$
		\ENDIF
	\ENDIF
\ENDFOR

\end{algorithmic}
\end{algorithm}

\begin{algorithm}[!ht]

\caption{The outer hull extraction algorithm for an arbitrary polyhedron.}
\label{al:polydec}

\begin{algorithmic}[1]

\REQUIRE A polyhedron $|M|$
\ENSURE $M_{solid}$ and $M_{sheet}$

\STATE Determine and orient a seed triangle $t_0$
\STATE Extract the edge-connected component $C$ containing $t_0$
\STATE run Algorithm \ref{al:conncomp} on $C, t_0$ and add the result to $M_{solid}$ and $M_{sheet}$

\FORALL {edge-connected component $C$ in $M_{solid} \cup M_{sheet}$}
	\FORALL {edge-connected component $D$ in $M_{solid}$}
		\IF {$C$ is contained in the space bounded by $D$}
			\STATE Delete $C$ (from either $M_{solid}$ or $M_{sheet}$)
		\ENDIF
	\ENDFOR
\ENDFOR

\end{algorithmic}
\end{algorithm}

If the resulting $M_{sheet}$ is empty (i.e. the polyhedron has a closed outer hull), $M_{solid}$ is the union of manifold-connected printable components (def. \ref{def:printable}) and represents the exact outer hull of the input. Each single manifold-connected component can be extracted \cite{deflosgp06} and saved to a separate STL file.

If $M_{sheet}$ is not empty, we consider the minimum thickness $\epsilon$ that the target printer is able to actually build. Then, we take $M_{sheet}$, orient it in an arbitrary direction, and produce a copy $C'$ with an inverted orientation. $C$ and $C'$ are then stitched along their common boundary, and the resulting closed mesh is \emph{inflated} through an offsetting at distance $\epsilon/2$ \cite{qu03}.
Possible unorientable configurations are properly cut to make them orientable before producing $C'$.
The union of $M_{solid}$ and the so-inflated $M_{sheet}$ undergoes the whole repairing process once more (both Sect. \ref{sec:stltopoly} and Sect. \ref{sec:polytoprint}), but this time we are guaranteed that our algorithm will produce an empty $M_{sheet}$.

Overall, this algorithm guarantees that all the solid parts are fixed without any distortion, whereas a minimum modification is introduced to fix sheet-like parts.

For the sake of simplicity, the treatment of visible \emph{wire-like} features has been omitted. Though in principle an STL file should not represent isolated edges, however, wire-like features can be defined as either irregular triangles or triangles having a degenerate realization. To treat these cases, we store all the removed degenerate and irregular triangles as segments in a separated list $D_t$. After the execution of the main algorithm, each segment in $D_t$ is analyzed: if the entire segment is part of the surface of either $M_{solid}$ or $M_{sheet}$, or if it is entirely contained in the volume enclosed by $M_{solid}$, then it is discarded; otherwise it is \emph{inflated} into an $\epsilon$-diameter cylinder and undergoes the second iteration of the repairing process along with the inflated $M_{sheet}$.

\section{Results and discussion}
\label{sec:results}
Experiments were run on a 2.67 GHz Intel Core i7 PC with 6 Gb RAM. Our C++ implementation resolves self-intersections \cite{attene2014} and encodes the resulting non-manifold complex in a proper data structure \cite{deflo03} which forms the input to our algorithm.
A Flashforge Kreator 3D printer was used to fabricate the physical prototypes using standard ABS filament. Printing parameters were set by the Skeinforge slicer based on the default configuration for printers belonging to the Makerbot family.

\begin{figure*}[!tb]
    \centering
  \includegraphics[width=.99\linewidth]{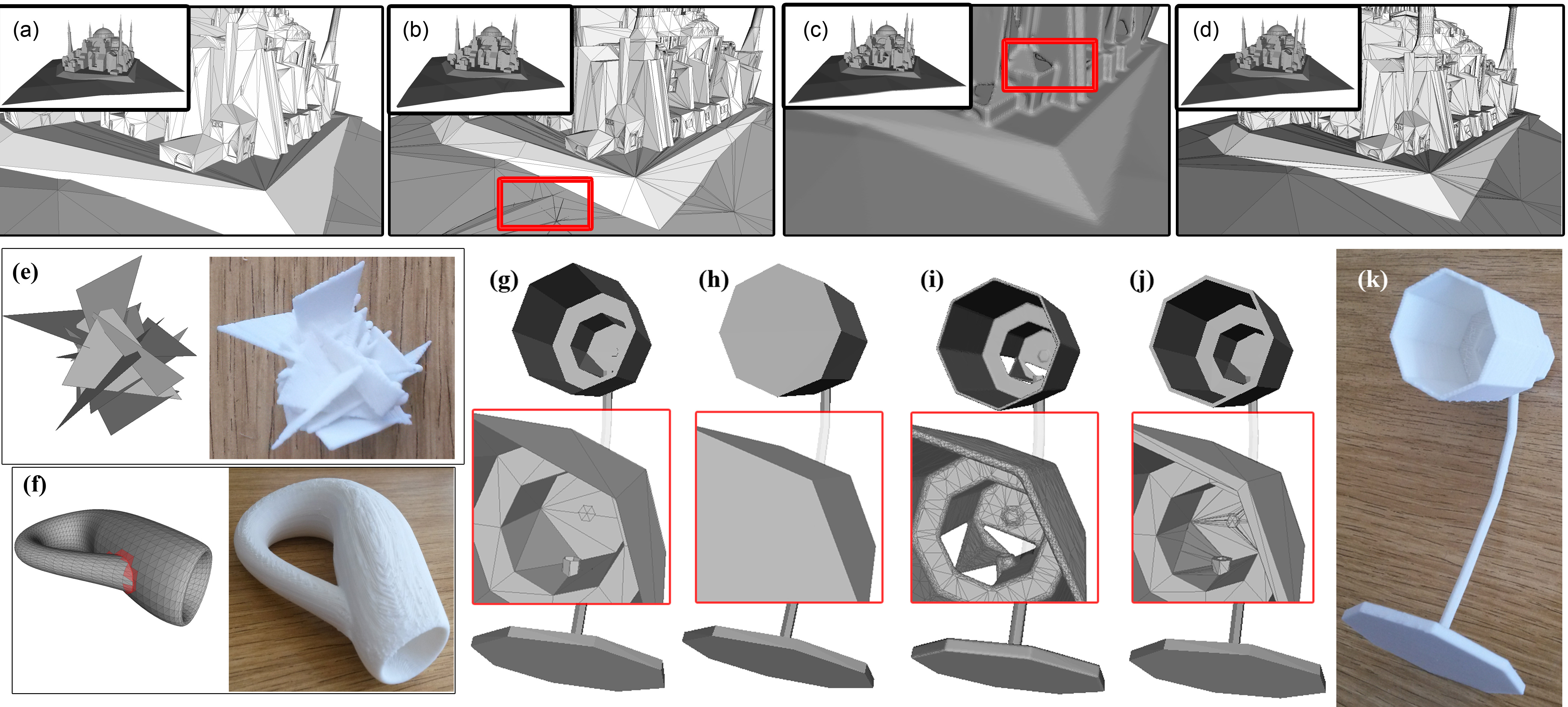}
  \caption{\label{fig:hagia}The original Hagia Sophia model (a) contains many heterogeneous defects, including self-intersections, unorientable surfaces, and zero-thickness patches (the ground). NetFABB result is still self-intersecting ((b), see red box), whereas Meshmixer introduced a significant distortion (c). In contrast, our algorithm could produce a printable model (d) whose visible surface is identical to the input with the exception of the thickened ground.
  Two challenging models such as the random triangles (e) and the Klein bottle ((f) with its necessarily self-intersecting embedding (red triangles)) could be successfully repaired by our method and then printed.
  The original Princeton lamp STL file with open surfaces used to model sheet-like features (g) and three repaired versions produced by NetFABB cloud service (h), Autodesk Meshmixer (i), and our algorithm (j). On the right (k), a physical replica of the model in (j) is shown.}
\end{figure*}

Experiments were run on all the 1814 models of the Princeton Shape Benchmark \cite{psb} and on particularly challenging additional models (Figs. \ref{fig:darth} and \ref{fig:hagia}). The prototype succeeded in all the cases with an average speed of 11K triangles per second. The speed depends on a number of factors, but the total elapsed time is largely dominated by the self-intersection removal phase: all the aspects that determine the efficiency for this phase are already described in \cite{attene2014} where results are reported for models made of up to millions of triangles. Herewith we just observe that the self-intersection removal must be run twice if $M_{sheet}$ is not empty, and the second iteration usually deals with more triangles. Hence, during the experimentation we also measured the average speed on models with empty $M_{sheet}$ (16K tri/sec) and on all the other models where two iterations were required (5K tri/sec). All the models were first scaled to have a bounding box maximum extension equal to 100 mm, and a value of 0.4 mm was used for $\epsilon$ (this is the extruder nozzle diameter of the 3D printer used for the experiments).

%

We compare against the most popular repairing systems for 3D printing applications, namely, NetFABB (https://netfabb.azurewebsites.net/) and Meshmixer (http://www.meshmixer.com). Our evaluation protocol rates each method according to the following scoring table:
\begin{itemize}
\item 1: Failure. Output is empty, unprintable, or completely different from the input (e.g. NetFABB creates a single tetrahedron out of the model in Figure \ref{fig:hagia}(e));
\item 2: Incomplete. Output is printable and resembles the input, but it misses parts which were visible in the input;
\item 3: Distorted. The output is printable and all the visible parts of the input are represented, but it is unnecessarily distorted;
\item 4: As-exact-as-possible. The output is printable, complete, and without unnecessary distortions (Sect. \ref{sec:terminology});
\item 5: Exact. The output is printable, complete, and without any distortion.
\end{itemize}

Based on such a protocol, the behavior of each algorithm could be rated on a per-model basis. As shown in Table \ref{tab:results}, our method outperforms the competing algorithms in most difficult cases.

\begin{table}
\centering
\begin{tabular}{ | l | l | c | c | c |}
\hline
\textbf{Model} 		&Figure	   &NetFABB &MeshMixer &ours\\
\hline \hline
Klein bottle	&\ref{fig:hagia}(f)	&5	&3	&5\\ \hline
Princeton Lamp	&\ref{fig:hagia}(g)		&2	&2	&4\\ \hline
Random Triangles &\ref{fig:hagia}(e)	&1	&3	&4\\ \hline
Darth Vader		&\ref{fig:darth}	&2	&3	&4\\ \hline
Hagia Sophia	&\ref{fig:hagia}(a)	&1	&2	&4\\ \hline
\end{tabular}
\caption{Results of our experiments. Scores: 1 = failure; 2 = incomplete; 3 = distorted; 4 = as-exact-as-possible; 5 = exact.}
\label{tab:results}
\end{table}


Unfortunately, NetFABB is run remotely on a proprietary hardware whose characteristics are not public. Hence, computational efficiency could be fairly compared against MeshMixer only. MeshMixer took 110 seconds to produce the fixed version of the Darth Vader model (15294 triangles, Figure \ref{fig:darth}), whereas our algorithm needed less than 5 seconds. Furthermore, it is worth mentioning that the output model produced by MeshMixer is made of more than one million triangles, whereas our result is much simpler (19541 triangles). A similar behavior could be observed on all the other models, and in general we calculated that our method is more than one order of magnitude faster than MeshMixer while producing more accurate and lightweight results (See table \ref{tab:results_nums}).
We point out that the triangle count is an important aspect in 3D printing applications because the slicing software can require quite long elaborations on too complex models. For example, the default slicer distributed with our 3D printer (i.e. Skeinforge) needs 13 minutes to process the Darth Vader model fixed by MeshMixer, while 94 seconds were sufficient to slice our output.
It is worth specifying that MeshMixer converts the input model into a volumetric representation and generates the output by reconstructing a mesh out of it. Hence, the repairing speed and eventual triangle count depend on the sampling density. Apparently MeshMixer does not allow to specify such a density explicitly, but even with different densities a fundamental problem would remain: the underlying approach which rebuilds a new mesh from a volumetric description, indeed, leads to an approximation of the input model. The resulting mesh can be made simpler by reducing the sampling density, but this would lead to a further loss of accuracy, while our objective is to be as exact as possible. The same argument holds if mesh simplification algorithms are used to reduce the triangle count before the slicing phase. Furthermore, traditional mesh simplification \cite{garlandqem} can easily produce results with self-intersections. This happens, for example, when simplifying the 1M triangles Darth Vader model produced by MeshMixer using the "Quadric Edge Collapse Decimation" function in MeshLab \cite{meshlab}.

\begin{table*}
\centering
\begin{tabular}{ | l | c | c | c | c | c | c | c | c | c | c | c | c | c | c | c |}
\hline
\textbf{M} 		&Orig. & &	   &NetFABB  & & & &MeshMixer & & & &ours & & &\\
		&\#T &\#KB &\#I		&\#T &\#KB &\#I &D		&\#T &\#KB &\#I &D		&\#T &\#KB &\#I &D\\
\hline \hline
1	&4002 &781 &98 &4564 &890 &296 &0 &32738 &6387 &0 &0.43 &4394 &857 &0 &0.0\\ \hline
2	&145 &28 &42 &264 &50 &0 &11.3 &30678 &5981 &0 &0.47 &688 &131 &0 &0.40\\ \hline
3 &20 &4 &18 &12 &3 &0 &4.9 &26122 &4876 &0 &0.41 &5358 &1013 &0 &0.41\\ \hline
4		&15294 &4176 &5060 &21064 &4653 &1102 &18.5 &1058720 &223453 &0 &0.46 &88442 &19541 &0 &0.42\\ \hline
5	&52851 &11842 &7914 &59712 &13378 &2142 &3.8 &1337174 &302297 &0 &0.45 &144046 &32298 &0 &0.41\\ \hline
\end{tabular}
\caption{This table reports our measurements on the original model and on the three fixed versions produced by NetFABB, MeshMixer and our method respectively. Legenda: M = Model ID (1 = Klein bottle, 2 = Princeton Lamp, 3 = Random triangles, 4 = Darth Vader, 5 = Hagia Sophia); \#T = number of triangles; \#KB = ASCII STL File size in KBytes; \#I = number of intersecting triangles; D = maximum deviation of the fixed model wrt the input STL (in mm). The maximum deviation introduced by MeshMixer is comparable to ours. However, while this deviation is more or less constant throughout the whole surface for MeshMixer, it is concentrated only over the sheets in our method. Indeed, the value of D in our method is zero for the Klein bottle because this model has no sheets.}
\label{tab:results_nums}
\end{table*}

\subsection{Discussion}
Besides computational efficiency, we observe that NetFABB detects boundaries in the input abstract complex and, if any, patches them before running the repairing. Besides providing rough approximations (e.g. Figs. \ref{fig:darth} and \ref{fig:hagia}(g)), this approach is not guaranteed to generate a solid in all the cases because the geometric realization may have zero thickness.

\begin{figure}[ht]
    \centering
  \includegraphics[width=.80\linewidth]{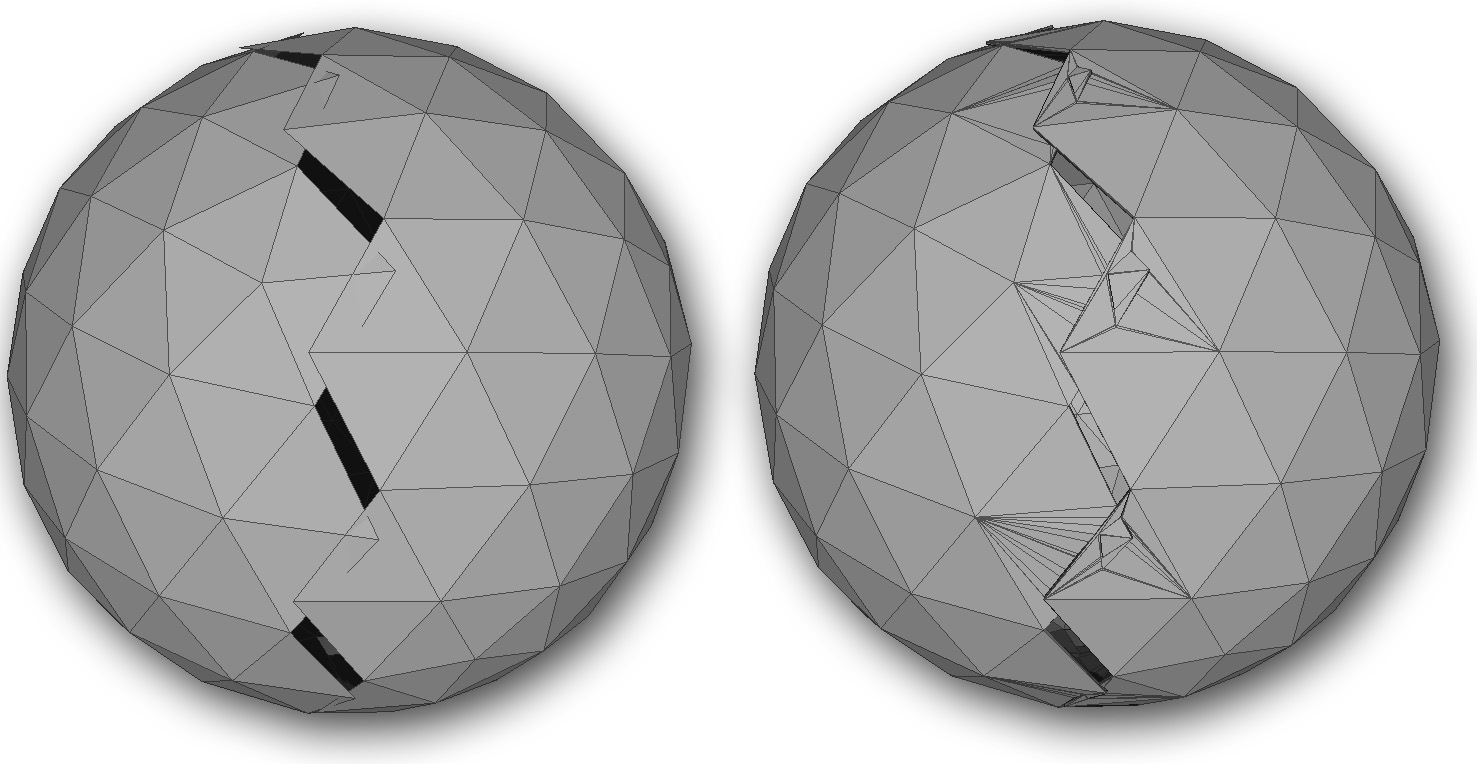}
  \caption{\label{fig:brokensphere}An original spherical model made of two inaccurately tessellated patches ((a), two connected components, two boundary loops, no topological handles), and its repaired version ((b), one connected component, zero boundary loops, 11 handles). Though the repaired model is printable, the actual design intention was probably different (one connected component, zero boundary loops and zero handles).}
\end{figure}

In principle, sheet-like parts in the outer hull can be either designed on purpose or produced by inaccurate processes. In the former case, for example, a non-expert designer may have idealized a thin part through a zero-thickness surface (see for example Figure \ref{fig:hagia}(g)). Conversely, as an example of the latter case we can simply consider the tessellation of a sphere composed of adjacent surfaces: without particular care, the piecewise linear approximations of the borders of adjacent patches may not match exactly \cite{bischoff05} and leave open holes and gaps.
We recognize that the repairing procedure described so far can produce an unexpected result when a sheet-like part is due to inaccuracy (e.g. Figure \ref{fig:brokensphere}), but we also observe that an as-exact-as-possible method is mostly necessary when the input must be precisely preserved.
After all, providing a completely automatic algorithm appears to be very important for a 3D printing community where many designers do not have a background in surface topology, but in a client application an expert user can still be allowed to interact with the process to distinguish inaccuracies from intentionally-designed features so that, e.g., a surface hole can be patched in advance or at the most appropriate point along the process.


Also, we point out that in this context the value of $\epsilon$ reflects the actual capability of the printer to produce a thin part (e.g. the diameter of the extruder nozzle) and is not meant to account for the structural strength of the physical prototype. To cope with these issues, our solution can be followed by algorithms such as \cite{stava2012}. 

One might argue that all the machinery within Algorithm \ref{al:polydec} is not really necessary because, in principle, we could simply construct a constrained Delaunay tetrahedrization (CDT) of the non-manifold complex and remove the outer tets up to constrained triangles. Actually, this solution is not that easy to implement for several reasons. First, to resolve the self-intersections without the risk of inserting new flaws it can be necessary to rely on exact arithmetic to represent coordinates \cite{attene2014}, and efficient CDT algorithms such as TetGEN \cite{si2005} do not support this input. On the other hand, though appropriate tools exist \cite{fabri2009cgal}, computing the CDT on exactly represented coordinates appears to be a too slow and memory demanding operation, even without counting that the skinny triangles which are typically produced to cut intersecting meshes would require the use of numerous Steiner points.

Note that the Topological Filtering introduced in \cite{attene2014} can produce similar results when $M_{sheet}$ is guaranteed to be empty; however, in this previous work the triangle orientation provided by the input STL is considered reliable enough to produce topological changes, and though this can be considered a positive aspect in some contexts, unorientable surfaces cannot be resolved by the Topological Filtering approach. Conversely, our variation covers all the cases where a closed outer hull is well-defined, independently of the orientability (e.g. Figure \ref{fig:hagia}(f)).

Finally, the model might be perfectly well defined with neither representational nor design defects, but it can still be not suitable for printing due to an incompatibility with the specific printer to be used (e.g. too thin walls, too tiny features, size larger than printing volume). We will not deal with these issues here, but existing works tackle the problem for specific printing devices and scenarios \cite{pintus2010} \cite{luo2012}.

\subsection{User interaction}
The repairing algorithm described in this paper can be part of more generic systems which involve the user in the ill-posed task of distinguishing between intentional thin sheets and unintentional disconnections in the outer surface.
Such a system may run the repairing process once and, before thickening $M_{sheet}$, may use a threshold distance $\gamma > \epsilon$ to determine which portions of the boundary could be \emph{stitched} to other parts of the surface. At this point, the user is called into play to select which of these potential stitchings should actually take place: if the user selects at least one such stitching, the geometry is modified accordingly and the whole repairing is run once again.

\begin{figure}[!tb]
    \centering
  \includegraphics[width=.99\linewidth]{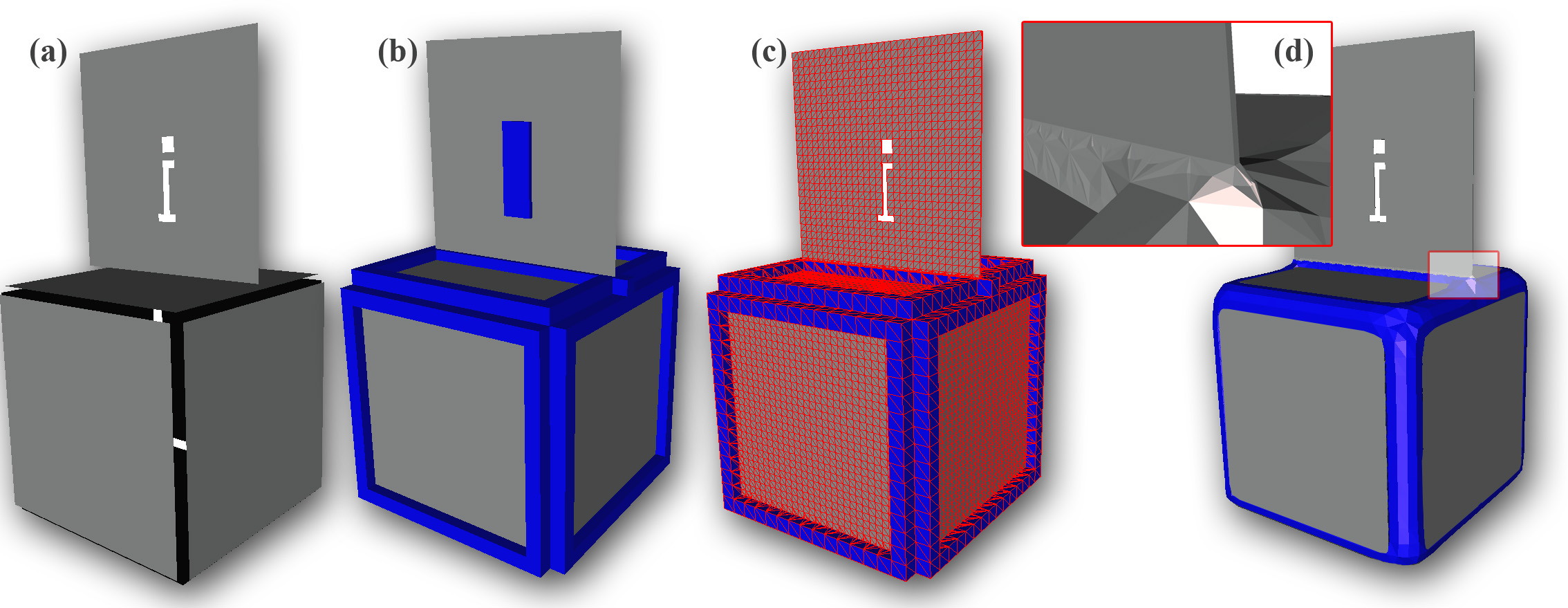}
  \caption{\label{fig:reconstruction}A model made of seven disconnected sheets with some unintended boundaries (a) is repaired and partitioned into $M_{solid}$ and $M_{sheet}$. The potential stitching areas are shown as voxel collections around the boundaries (b). After the user selection, the (triangulated) stitching areas are considered to be part of the input, and the repairing is re-run on such an integrated input: during this last repairing step, the algorithm keeps track of the triangles which were part of the stitching areas and keeps them selected (c). Vertices which are in the interior of a selected area are iteratively moved towards their centers of mass, so as to perform a local Laplacian smoothing (d). The resulting model is a single solid with two through holes representing the 'i' character.}
\end{figure}

To demonstrate this approach, a prototype interactive system has been implemented so that inaccurately-modeled objects (e.g. Figure \ref{fig:brokensphere}) could be repaired as expected. The prototype exploits a stitching procedure inspired on \cite{bischoff05}, and works as follows: the input STL undergoes the repairing process up to the first iteration of Algorithm \ref{al:polydec}. If $M_{sheet}$ is empty we already have the result and no ambiguity requires user interaction. Otherwise, the user is warned and asked to set a value for $\gamma$. A uniform grid is created to intersect model with cubical cells of size $\gamma / 2$. Each cell $C$ which is intersected by at least one boundary curve is analyzed: if the mesh restricted to the volume of $C$ is disconnected (i.e. $(M_{solid} \cup M_{sheet}) \cap C$ is disconnected), then $C$ is marked as \emph{potential stitch}; if not, the same check is repeated on each of the eight blocks of cells that share one of $C$'s vertices; specifically, if $v$ is one of the eight vertices of $C$, we consider the block $C'$ made of the eight cells incident at $v$, and if the mesh restricted to $C'$ is disconnected, we mark all the eight cells in $C'$ as \emph{potential stitch}.
The collection of all the potential stitches is displayed, and the user selects those that must lead to an actual stitch. These selected regions are triangulated, that is, each square forming the boundary of the region is split into two triangles. These new triangles are added to the mesh and the repairing is launched once again, but this time the thickening process and the second iteration are included. During this last repairing step, however, triangles which are part of the stitching areas are kept selected, and their subtriangles (i.e. those triangles that are generated by resolving their intersections) inherit such a selection. In a final step, the selected areas are \emph{smoothed} by iteratively moving their internal vertices towards their respective centers of mass. This process is summarized in Figure \ref{fig:reconstruction}.
It is important to consider that this example system contains the main algorithm but is not as-exact-as-possible. However, it shows how easily the main algorithm can be customized to implement extremely powerful and flexible repairing systems.

\subsection{Limitations}
Our algorithm is not meant to treat models with internal cavities. Surfaces that bound cavities, indeed, are not reachable from infinity, and therefore are not part of the outer hull according to our definition. Our definition can probably be extended to include these cases as well, but many additive manufacturing technologies cannot be used in any case to create such models because support material would remain trapped within the cavities. 
Furthermore, independently of the printing technology, we assume that the input STL is a raw representation of the object to be manufactured. Based on this, we can properly define the outer hull even if the orientation of triangles is unreliable. On the other hand, this generality makes us unable to distinguish an actual cavity from a topological artefact in the input.
Hence, extending our definition to an \emph{outer and inner hull} which includes cavities seems to be rather difficult: it is easy to select a point at "infinity" to define the entire outer hull, but it is not clear how one should define the cavities while assuming an unreliable orientation of the triangles. If the input can be guaranteed to have reliably oriented triangles, such a definition may take advantage from the notion of generalized winding numbers \cite{jacobson13}. In all the other cases, one may think of involving the user in the disambiguation process: by providing a point which is known to be part of a cavity, the cavity itself can be defined as in Def. \ref{def:outerhull} while considering this point instead of $p$. Any of these solutions, however, would require to adapt Algorithm \ref{al:polydec}.

\section{Conclusion and future work}
We have formally defined the class of printable STL files and have shown how to convert a generic STL to a printable model with no or minimum distortion. In particular, we have shown that the solid parts of the input can be fixed with no visible deformations, whereas zero-thickness surfaces must be necessarily made solid to become printable, and hence visible in the eventual physical prototype.

An interesting objective for future research is the automatic distinction of intentionally designed thin features and unintentional open surfaces due to inaccurate modeling. A potential inspiration here may come from the notion of generalized winding numbers \cite{jacobson13}, though this method cannot be readily exploited because it heavily relies on the triangle orientation. Thresholds and reasoning on distance fields can be alternative starting points.

\bibliographystyle{acmsiggraph}
\bibliography{printable_shapes}

\end{document}